\documentclass[aps,english,preprintnumbers,amsmath,amssymb,nofootinbib,twocolumn,superscriptaddress]{revtex4-1}

\pdfoutput=1

\usepackage[latin1]{inputenc}
\usepackage{graphicx}
\usepackage{bbm}
\usepackage{amssymb}
\usepackage{amsmath}
\usepackage{tabularx}
\usepackage{MnSymbol}
\usepackage{bigints}

\usepackage{color}
\usepackage{csquotes}

\usepackage{dsfont}

\def\0#1#2{\frac{#1}{#2}}

\def\s0#1#2{\mbox{\small{$ \frac{#1}{#2} $}}}

\newcommand{\be}{\begin{eqnarray}}
\newcommand{\ee}{\end{eqnarray}}

\newcommand{\nn}{\nonumber }

\newcommand{\beq}{\begin{equation}}
\newcommand{\eeq}{\end{equation}}
\newcommand{\bea}{\begin{eqnarray}}
\newcommand{\eea}{\end{eqnarray}}

\newcommand{\m}[1]{\ensuremath{\mathrm{#1}}}

\usepackage{babel}
\makeatother
\begin{document}

\title{In-medium bound-state formation and inhomogeneous condensation\\ in Fermi gases in 
a hard-wall box}

\author{Dietrich Roscher}
\affiliation{Institut f\"ur Theoretische Physik, Universit\"at zu K\"oln, D-50937 Cologne, Germany}
\affiliation{Institut f\"ur Kernphysik (Theoriezentrum), Technische Universit\"at Darmstadt, 
D-64289 Darmstadt, Germany}
\author{Jens Braun}
\affiliation{Institut f\"ur Kernphysik (Theoriezentrum), Technische Universit\"at Darmstadt, 
D-64289 Darmstadt, Germany}
\affiliation{ExtreMe Matter Institute EMMI, GSI, Planckstra{\ss}e 1, D-64291 Darmstadt, Germany}

\begin{abstract}
The formation of bosonic bound states underlies the formation of a superfluid ground state in the many-body phase diagram of ultracold
Fermi gases. We study bound-state formation in a spin- and mass-imbalanced ultracold Fermi gas confined in a box with hard-wall boundary
conditions. Because of the presence of finite Fermi spheres, the center-of-mass
momentum of the potentially formed bound states can be finite, depending on the parameters controlling mass and spin imbalance
as well as the coupling strength. We exploit this observation to estimate the potential location of inhomogeneous phases in the 
many-body phase diagram as a function of spin- and mass imbalance as well as the box size. Our results suggest
that a hard-wall box does not alter substantially the many-body phase diagram calculated 
in the thermodynamic limit. Therefore, such a box may serve as an ideal trap potential to bring 
experiment and theory closely together and facilitate the search for exotic inhomogeneous ground states.
\end{abstract}

\maketitle

%
\section{Introduction}
The search for ground states associated with a spontaneous breakdown of translation
invariance in quantum many-body systems has inspired both experimental
and theoretical studies since the potential existence of such phases has been predicted independently by
{\it Fulde} and {\it Ferrell} as well as {\it Larkin} and {\it Ovchinnikov} in their seminal works~\cite{FuldeFerrell64,LarkinOvchinnikov64}.
Loosely speaking, the formation of such so-called FFLO-type ground states described by a spatially varying condensate
in fermionic theories is directly related to the formation of bosonic two-fermion bound states. The macroscopic
occupation of the energetically lowest-lying bosonic state is then associated with the spontaneous breakdown of a 
fundamental symmetry of the underlying fermionic theory and
the emergence of a condensate. Depending on the control parameters in the system,
this lowest-lying bosonic state may carry a finite center-of-mass momentum and the macroscopic occupation 
of this state may then lead to the formation of a spatially varying ground state which is energetically
favored over the formation of a homogeneous ground state. However, 
a clean experimental verification of the existence of such inhomogeneous phases
is challenging since the presence of other length scales -- not present in theoretical studies in the thermodynamic limit 
but unavoidably present in an experimental setup -- may distort the system such that theses phase 
are no longer energetically favored or that at least the experimental signatures for the existence of these phases are 
strongly contaminated.

At this point, ultracold atomic Fermi gases come into play. Indeed, since the first realization of a {\it Bose-Einstein}-condensate 
of paired fermions~\cite{PhysRevLett.92.040403, Jochim_etal_FermBEC03}, experimental techniques
have been pushed to an unprecedented level and by now provide an 
accessible, versatile and very clean environment to study quantum many-body phenomena, ranging from 
{\it Bose-Einstein} condensation (BEC) to {\it Bardeen-Cooper-Schrieffer} (BCS) superfluidity,
including a control over temperature 
and polarization~\cite{Zwierlein27012006, *Partridge27012006,*2006Natur.442...54Z, *PhysRevLett.97.030401, *PhysRevLett.97.190407, *Schunck11052007, *2008Natur.451..689S},
see Refs.~\cite{ketterle-review,Bloch:2008zzb,Giorgini:2008zz} for reviews. 
More recent developments now even open up the possibility to study mixtures of a variety of different fermion species (e.g. ${}^6$Li, ${}^{40}$K, ${}^{161}$Dy, ${}^{163}$Dy, and ${}^{167}$Er), 
see e.g. Refs.~\cite{GrimmPC,2012PhRvL.108u5301L,2013PhRvA..88c2508F}.
Apart from their phenomenological relevance for our understanding of quantum many-body phenomena, 
these experiments have reached high precision in many cases such that they can be used to benchmark theoretical methods~\cite{Zwerger-book,ChevyMora,StoofGubbels}.
\begin{figure*}[t]
\centering
\includegraphics[width = 1.8\columnwidth]{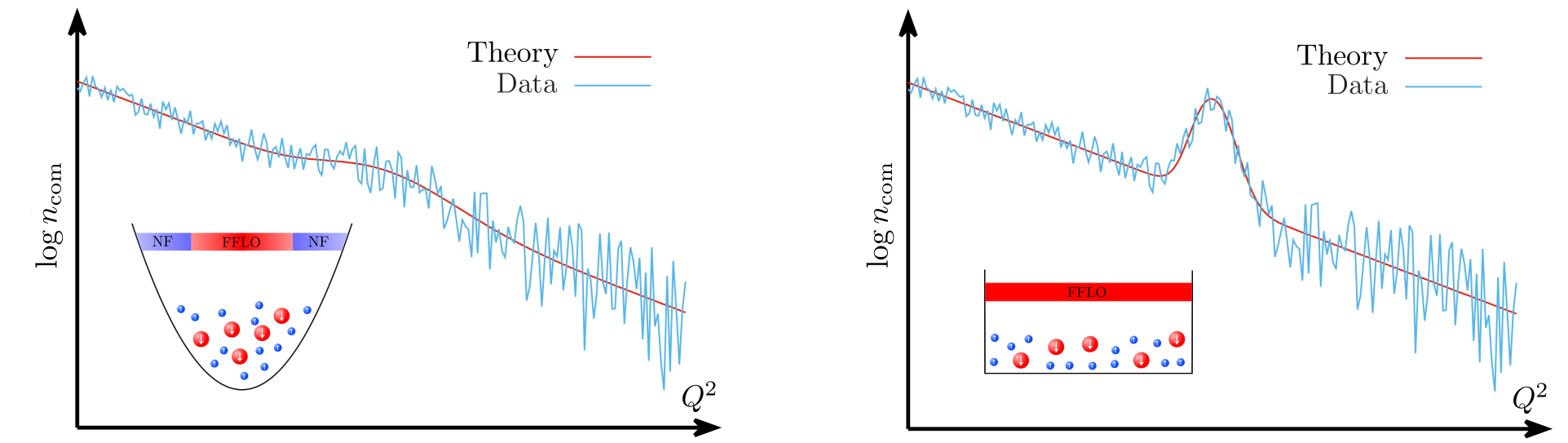}
\caption{Sketch of the finite-temperature center-of-mass momentum distribution of bound states 
formed in a Fermi gas confined in a one-dimensional harmonic trap (left panel) and in a hard-wall box (right panel). 
The potential formation of an FFLO-type ground state is associated with a maximum located at a finite momentum.}
\label{fig:momdist}
\end{figure*}

In experiments with ultracold atomic Fermi gases, the particle density $n$ (i.e. the Fermi momentum~$k_{\rm F}$) and the 
$s$-wave scattering length $a_s$ are control parameters. 
At least for a sufficiently dilute gas in the absence of a confining trap potential,
they even represent the only scales of the system
since the effective range of the interaction can safely be neglected. 
Experimentally, $a_s$ can be tuned by means of so-called magnetic {\it Feshbach} resonances. 
This opens up the possibility to explore many-body phenomena over a wide range of interaction strengths since
the associated coupling~$g$ of the underlying contact interaction is directly related to the $s$-wave scattering length, e.g.~$g\sim 1/a_s$ 
in one dimension~\cite{2000EJPh21.435B}. 

For a study of the existence of FFLO-type phases, 
the control over polarization and/or mass imbalance in experiments with two-component ultracold Fermi gases is 
crucial. Indeed, the emergence of such phases is intimately related
to the sizes of the Fermi surfaces associated with the two components~\cite{FuldeFerrell64,LarkinOvchinnikov64}.
The latter can be changed relative to each other by varying the polarization or by studying fermions with different masses. 
A significant mismatch in the sizes of the Fermi surfaces is then expected to trigger the formation of an inhomogeneous
condensate. For three-dimensional spin-imbalanced Fermi gases, however, theoretical studies even beyond
the mean-field approximation suggest that
inhomogeneous condensation is only favored in a thin layer as a function of the spin-polarization, 
if at all~\cite{SheehyRadzihovskyPRL06,HuLiuMFh06,Bulgac:2008tm,Boettcher:2014tfa,Roscher:2015xha},
see also Ref.~\cite{Radzihovsky2012189} for a review.
Although the extent of the inhomogeneous phase in parameter space has been found to increase significantly 
when the mass imbalance is increased~\cite{Wang_etal_mIFFLO14,Roscher:2015xha}, its size is still rather small and 
therefore the detection of such a phase would still represent a major challenge in future experiments with
mass-imbalanced Fermi gases. On the other hand, theoretical studies suggest that the phase diagram of a spin-imbalanced one-dimensional
Fermi gas is to a large extent occupied by an FFLO-type 
phase~\cite{PhysRevB.63.140511,PhysRevLett.98.070402,PhysRevLett.98.070403,PhysRevLett.99.250403,PhysRevA.78.033607,Roscher:2013cma} and 
indications for the existence of this phase have been indeed found in experiments in tightly constraining trap potentials~\cite{Liao_etal_1DFFLO10}.  

Regarding the detection of inhomogeneous condensation, 
an intriguing relation between the general structure of the many-body phase diagram and the center-of-mass momentum
of the bound-states formed in a spin- and mass-imbalanced medium 
has been observed in one and three dimensions~\cite{Roscher:2013cma,Roscher:2015xha}. 
In fact, an inhomogeneous phase has only been found to be favored in those regimes of the spin- and mass-imbalanced
many-body system where the corresponding study of the in-medium two-body problem indicates that the formation of 
a bound state with a finite center-of-mass momentum is energetically favored. This observation sets the stage for
our present work. 

Despite the fact that the size of the inhomogeneous phase in parameter space might be small, the experimental 
search for inhomogeneous phases is further complicated by the presence of a trap potential which, in case of, e.g., a harmonic trap, introduces
an explicit inhomogeneity which distorts the Fermi surfaces by rendering them effectively space-dependent. The consequences
of effectively space-dependent Fermi surfaces are best illustrated with the aid of Density Functional
Theory (DFT) in the Local Density Approximation (LDA) which relies on the continuum equation
of state of the spin- and mass-imbalanced Fermi gas as 
an input, see, e.g., Refs.~\cite{Bulgac:2007ah,PhysRevA.75.023610,RLS,KBS,2010Natur.467..567L,Braun:2014ewa}.
To this end, we first note that the Fermi surfaces can be related to the chemical potentials of the two fermion
components. Space-dependent chemical potentials then imply that the underlying continuum equation
of state is effectively probed at different points in parameter space, depending on the actual
distance from the center of the trap potential. This is already sufficient to 
understand on a qualitative level the shell structure of ultracold Fermi gases in harmonic traps 
observed in experiments~\cite{2006PhRvL..97c0401S,PhysRevLett.97.030401,Partridge27012006,PhysRevLett.97.190407,2010Natur.467..567L}.
Since many-body calculations in the thermodynamic limit find that the extent of the inhomogeneous phase as a function of spin
polarization is comparatively small~\cite{SheehyRadzihovskyPRL06,HuLiuMFh06,Bulgac:2008tm,Boettcher:2014tfa,Wang_etal_mIFFLO14,Roscher:2015xha},
even for finite mass imbalances, an inhomogeneous phase is only expected to appear in a thin layer around the superfluid
homogeneous core in, e.g., a three-dimensional harmonically trapped Fermi gas. 

In this work, we are eventually interested in the formation of an inhomogeneous ground state. As this phenomenon
is intimately related to the existence of bound states, the bound-state momentum distribution is of particular interest. 
For example, the formation of a homogeneous condensate is associated with a significant
increase of the value of this momentum distribution at vanishing momentum.
In turn, inhomogeneous condensation would be indicated by a significant increase in the population
of states with a finite center-of-mass momentum resulting in a maximum of the momentum
distribution at a finite momentum. Following up on our line of 
arguments based on DFT in LDA, however, it is then reasonable to expect that inhomogeneous condensation in a harmonically trapped gas 
will be indicated by a rather broad maximum in the momentum distribution, if resolvable at all. Indeed,
bound states with center-of-mass momenta spread over a wide range may contribute to
the formation of the inhomogeneous ground state due to the effective space dependence
of the chemical potentials, see left panel of Fig.~\ref{fig:momdist} for an illustration.
The situation may change drastically when we consider a box with hard-wall boundary 
conditions as a trap potential, see Refs.~\cite{2016arXiv161010100M,PhysRevLett.110.200406} for first experimental explorations. 
On the theory side, {exact solutions of the two-body vacuum problem in a harmonic trap potential~\cite{Busch1998}, 
periodic box~\cite{BraunPB}, and hard-wall box~\cite{Belloni201425} have been found. 
However, many-body effects in systems with hard-wall boundary
conditions have been largely unexplored, mostly related 
to an increase of the complexity, e.g., associated with the explicit breaking of translation invariance.
The hard-wall potential is constant between the}
confining hard walls and therefore also the chemical potentials remain constant
within this type of a trap, as it is in the thermodynamic limit. Leaving boundary 
effects close to the hard walls aside, which are only expected to be prominent in the limit
of a small number of fermions, we expect that no distinct shell structure 
emerges in this case and a potentially existing maximum in the 
momentum distribution should be more pronounced as in the case of 
a harmonic trap, see right panel of Fig.~\ref{fig:momdist} for an illustration. These 
considerations represent the basis for our analysis of particle-number and 
finite-size effects in ultracold Fermi gases confined in a hard-wall box.

In the following we focus on one-dimensional systems in
our explicit calculations for simplicity. In general, studies of a macroscopic 
occupation of the ground state as associated with the formation of a condensate 
is delicate in one-dimensional systems since long-range fluctuations hinder the 
spontaneous breakdown of a continuous symmetry in this case~\cite{PhysRevLett.17.1133,PhysRev.158.383}.
With respect to the conclusions drawn from our present in-medium two-body calculations for the many-body phase diagram, 
however, we restrict ourselves to studies of gases in a hard-wall box. Here, the box 
acts as a physical infrared cutoff. Condensation associated with a macroscopic occupation of the ground state 
is then possible even in one dimension 
since long-range fluctuations are cut off by the confining potential,
even in the non-interacting limit~\cite{PhysRevA.54.656}. In the infinite-volume limit, this condensate will
eventually fade away, as follows from general considerations~\cite{PhysRevLett.17.1133,PhysRev.158.383}. On the
other hand, the appearance
of a condensate in the presence of a finite infrared cutoff is closely related to the existence of
a precondensation phase in three-dimensional gases, where 
bound states are formed and local phase coherence exists although the system is 
in the normal phase with no long-range order due to fluctuation effects~\cite{Boettcher201263,Boettcher:2014tfa}.
An analogous so-called local ordering phenomenon has also been found to exist
in relativistic theories, see, e.g., Ref.~\cite{Braun:2009si}. We add that precondensation  is  closely  
related to the so-called pseudogap phase where low-lying fermionic  excitations  are  gapped
but the many-body system is in the normal phase~\cite{2010NatPh...6..569G,PhysRevLett.106.060402,PhysRevA.83.053623,PhysRevLett.114.075301}.

The rest of the paper is organized as follows: In Sec.~\ref{sec:form}, we introduce the formalism
underlying our study of in-medium bound-state formation in a hard-wall box. Our results 
are discussed in Sec.~\ref{sec:res}, including a discussion 
of the consequences for the many-body phase diagram. Our conclusions, also with respect to three-dimensional systems, 
are given in Sec.~\ref{sec:conc}.

\section{Formalism}\label{sec:form}

The formation of a macroscopically occupied ground state associated with a condensate of paired fermions
requires formation of these pairs in the first place. Indeed, studies of in-medium bound-state formation have
shown even semi-quantitative agreement with calculations of the many-body phase 
diagram in the thermodynamic limit~\cite{Roscher:2013cma,Roscher:2015xha}. In particular, the formation of bound states
with a finite center-of-mass momentum is found to be tightly connected to the formation of an inhomogeneous 
ground state associated with spontaneous breaking of translation invariance. 
Of course, the exact functional form associated with the inhomogeneity in the many-body
system cannot be predicted from a bare study of bound-state formation. 
However, the center-of-mass momentum of the formed bound states has
been found to be at least a good measure for the characteristic momentum scale associated with the inhomogeneity 
in many-body calculations in the thermodynamic limit~\cite{Roscher:2013cma,Roscher:2015xha}. For example, the inhomogeneity may be
described by a periodic function in position space where its frequency is effectively determined by the center-of-mass momentum
of the bound states.
Therefore, it is reasonable
to expect that the emergence of a macroscopically occupied ground state in a system of fermions confined
in a hard-wall box is also dominantly influenced by the properties of the energetically most favorable in-medium bound state. 

Following previous studies in the thermodynamic limit~\cite{Roscher:2013cma,Roscher:2015xha}, 
we now perform a calculation of the formation of bound states in a medium of fermions described
by inert\footnote{The Fermi seas entering our study are not fully inert, see Eq.~\eqref{eq:Seq}.
The dispersion relations of the two interacting fermions implicitly 
permit to excite fermions from the Fermi seas. Contrary to that, truly inert Fermi seas
would rather correspond to a situation where the dispersion relations of the (interacting) fermions 
in the {\it Schr\"odinger} equation are of the form~$\sim p^2\theta( p^2 - (2m_{\sigma})\epsilon_{{\rm F},\sigma})$.}
Fermi seas in a one-dimensional hard-wall box.
The {\it Schr\"odinger} equation for the wave function $\Psi$ of two distinct fermions 
in the presence of their respective Fermi seas reads\footnote{Note that the general setup has been originally used to determine the 
properties of a single Cooper pair in the context of balanced BCS theory, see, e.g., Refs.~\cite{Cooper56, PitaevskiiBook}.}
\be
\!\!\!\!\!\left[\epsilon_\uparrow(\partial_{x_\uparrow}) + \epsilon_\downarrow(\partial_{x_\downarrow}) - 
g\delta(x_\uparrow - x_\downarrow) + E_{\rm{B}}\right] \Psi(x_\uparrow,x_\downarrow) = 0\,,
\label{eq:Seq}
\ee
where~$E_{B}=\epsilon_{{\rm F},\uparrow}+ \epsilon_{{\rm F},\downarrow}-E$ is the energy of the bound state 
in the medium.
In our case, the interaction between the spin-up and spin-down fermion is given by a contact interaction.
The associated coupling strength~$g$ is determined by the $s$-wave scattering length~$a_s$, i.e. $g=1/a_s$~\cite{2000EJPh21.435B}. 
The kinetic energy of the fermions is measured with respect to their Fermi energies~$\epsilon_{{\rm F},\sigma}$: 
$\epsilon_\sigma(\partial_{x_\sigma}) = |-(2m_\sigma)^{-1}\partial^2_{x_\sigma} - \epsilon_{{\rm F},\sigma}|$,
where $\sigma=\{\uparrow,\downarrow\}$ and~$m_{\sigma}$
determines the masses of the respective species.
Since we consider a box with hard walls and extent~$L$ (i.e. an infinite potential well), 
the wave function~$\Psi(x_{\uparrow},x_{\downarrow})$ is finite only for $-L/2 < x_{\uparrow,\downarrow} < L/2$ and vanishes
identically otherwise. The complete orthonormalized set of eigenfunctions~$\{\varphi_k(x)\}$ for a 
single particle in such a potential is given by
\be
\varphi_k(x_{\sigma})=\sqrt{\frac{2}{L}}\sin\left(\frac{k\pi x_{\sigma}}{L} - \frac{k\pi}{2}\right)\,,\label{eq:ons}
\ee
which implies $\epsilon_{{\rm F},\sigma}=(2m_{\sigma})^{-1}(\pi (N_{\sigma}-1)/L)^2$ for the Fermi energies.
Note that the total number $N=N_{\uparrow}+N_{\downarrow}$ of spin-up and spin-down fermions includes
the two interacting fermions in our conventions. 

In order to solve the {\it Schr\"odinger} equation~\eqref{eq:Seq} for the ground state, 
we span the ground-state wave function~$\Psi(x_\uparrow,x_\downarrow)$ 
by the single-particle wave functions~$\{\varphi_k(x)\}$ defined in Eq.~\eqref{eq:ons} 
which already respect the boundary conditions set by the hard-wall box:
\be
\Psi(x_{\uparrow},x_{\downarrow}) = \sum_{k=1}^{N_{\rm B}}\sum_{l=1}^{N_{\rm B}}c_{kl}\varphi_k(x_{\uparrow})\varphi_l(x_{\downarrow})\,.
\label{eq:PsiExp}
\ee
The parameter~$N_{\rm B}$ specifies the size of our truncated basis set, 
see our discussion in Sec.~\ref{sec:res}. 
In the present work, the coefficients~$c_{kl}$ 
are obtained from a (numerical) diagonalization of the {\it Schr\"odinger} equation~\eqref{eq:Seq} and are associated
with the ground-state energy~$E_{\rm B}$, i.e. the lowest-lying state in our model.\footnote{Note that a solution in closed form 
of the {\it Schr\"odinger} equation~\eqref{eq:Seq} has been found in the thermodynamic limit~\cite{Roscher:2013cma,Roscher:2015xha}.}

A rigorous definition of a bound state is difficult in a hard-wall box as 
all states are bound by construction. 
Mathematically speaking, in the presence of a hard-wall box, the system is defined on
a compact support on which any continuous wave function permitted by the corresponding boundary conditions is square integrable.
To be consistent with studies in the thermodynamic limit\footnote{Here, 
thermodynamic limit refers to the limits~$N\to\infty$ and~$L\to\infty$ for a given 
fixed density of the fermions. In particular, we associate this limit as well as
the infinite-volume limit with a system in free space, i.e. a system in the absence of any kind of trap potential.
Note that, {\it a priori}, it is not clear that the approach of a given finite system to either
the thermodynamic or infinite-volume limit is continuous as~$L\to\infty$.},
we therefore define a state described by the wave function~$\Psi(x_{\uparrow},x_{\downarrow})$ to be 
a bound state if~$E_{\rm B}$ is positive\footnote{Recall our conventions for~$E_{\rm B}$
detailed below Eq.~\eqref{eq:Seq}.} 
for a given choice of the coupling~$g$ and 
the Fermi energies~$\epsilon_{{\rm F},\sigma}$, see Refs.~\cite{PitaevskiiBook,Roscher:2013cma,Roscher:2015xha}.

From the ground-state wave function~$\Psi$, we can compute the properties of our in-medium two-fermion system.
At this point, however, an important comment is in order: In a finite system with hard-wall boundary conditions,
the canonical momentum operator $\hat{p} = i\partial_x$ is {\it not} self-adjoint~\cite{2001AmJPh..69..322B,Belloni201425}. 
In particular, this implies that the center-of-mass momentum of a potentially
formed bound state is not a physical observable. To surmount this issue and extract information about the center-of-mass
momentum, being a key quantity in our analysis, we take the actual experimental setup into account. In experiments, the momenta of
the pairs are in general measured {\it in situ} right after the trap potential has been switched off. With respect to our study,
this renders the momentum a physical observable again and the bound-state momentum distribution can be
obtained from the Fourier transform of a quantity suitably constructed from the wave function~$\Psi$. 
To be specific, the center-of-mass (com) momentum distribution~$n_{\rm com}$ of the lowest-lying bosonic state (i.e. bound state, if~$E_{\rm B}>0$)
is given by
\be
n_{\rm com}(Q)= \bigintsss_{-\infty}^{\infty} \frac{{\rm d}q}{2\pi} \left|\Psi(q_{\uparrow}(q,Q),q_{\downarrow}(q,Q)) \right|^2\,.
\label{eq:ncom}
\ee
Here,~$q$ is the relative momentum of the two fermions and~$Q$ is their center-of-mass momentum. 
These quantities are related to the momenta~$q_{\uparrow}$ and~$q_{\downarrow}$ of the two fermions
via~$q_{\uparrow,\downarrow}(q,Q)=\frac{1}{2}(1\mp \bar{m})Q \pm q$, 
where the parameter~$\bar{m}=(m_{\downarrow}-m_{\uparrow})/(m_{\downarrow}+m_{\uparrow})$  
measures the relative mass difference of the two fermion species.
The ground-state wave function in momentum space~$\Psi(q_{\uparrow},q_{\downarrow})$ is defined as
\be
\!\!\!\Psi(q_{\uparrow},q_{\downarrow}) =  
\int_{-\infty}^{\infty}{\rm d}x_{\uparrow}\int_{-\infty}^{\infty}{\rm d}x_{\downarrow}\Psi(x_{\uparrow},x_{\downarrow})\,{\rm e}^{-{\rm i}(q_{\uparrow}x_{\uparrow} + q_{\downarrow}x_{\downarrow})}\,.
\ee
Since we have integrated over the relative momentum~$q$ of the two fermions in Eq.~\eqref{eq:ncom}, 
the distribution~$n_{\rm com}(Q)$ describes the probability density to find the two fermions with a center-of-mass
momentum~$Q$. {Thus, the position~$Q_0$ of the global maximum of~$n_{\rm com}$ determines the
most probable center-of-mass momentum of the two fermions.
If now~$Q_0\neq 0$ and the energy~$E_{\rm B}$ associated with the wave function~$\Psi$ is positive,
then the wave function~$\Psi$ is said to describe a bound state with a finite center-of-mass momentum~$Q_0$.}

In the infinite-volume limit, the wave function~$\Psi$
can be written as a product of a function describing the relative motion of the two fermions 
and another one describing their center-of-mass motion
~\cite{Roscher:2013cma,Roscher:2015xha}. In this case, we 
have~$n_{\rm com}(Q) \sim \delta(Q-Q_{0})+\delta(Q+Q_{0})$ with~$Q_0$ being the center-of-mass 
momentum of the bound state. Indeed, we observe in our numerical studies that the
center-of-mass momentum distribution develops a sharp maximum when the box size is increased,
see our discussion below.

If the lowest-lying state is indeed a bound state, then the formation of these states can be considered as a precursor of condensation.
A condensate of such bound states with a finite center-of-mass momentum would break translational invariance. 
Considering the parameter space spanned by the polarization~$\sim (N_{\uparrow}-N_{\downarrow})$
and the mass imbalance parameter~$\bar{m}$, the observation of a regime 
associated with the formation of bound states with a
finite center-of-mass momentum then suggests that the ground state of the many-body system
is potentially inhomogeneous for this set of parameters.

Before we present the results from our study of the in-medium two-body problem in a hard-wall box,
we finally discuss the distribution
function~$n_{\rm com}$ in the light of other distribution functions. For example, 
the so-called density-density correlation function is given by
\be
n_{n_{\uparrow}n_{\downarrow}}(x_{\uparrow},x_{\downarrow}) 
 =\langle \psi^{\dagger}_{\uparrow}(x_{\uparrow}) \psi_{\uparrow}(x_{\uparrow}) 
 \psi^{\dagger}_{\downarrow}(x_{\downarrow}) \psi_{\downarrow}(x_{\downarrow})\rangle\,,
\ee
where the operators~$\psi_{\sigma}^{(\dagger)}$ denote annihilation (creation) operators. 
In terms of a general $N$-body wave 
function~$\Phi(x_{\uparrow,1},x_{\downarrow,1},\dots,x_{\uparrow,N_{\uparrow}},x_{\downarrow,N_{\downarrow}})$, 
this correlation function can be written as follows:
\be
&&\!\!\!\!\!\! n_{n_{\uparrow}n_{\downarrow}}(x_{\uparrow},x_{\downarrow}) \nn\\
&& \;\; =N_{\uparrow}N_{\downarrow}\int_{-\infty}^{\infty}\!{\rm d}y_{3}\;\cdots\!\!\int_{-\infty}^{\infty}\! {\rm d}y_{N}|\Phi(x_{\uparrow},x_{\downarrow},y_3,\dots,y_N)|^2.
\ee
The in-medium two-body wave function~$\Psi$
describes the dynamics of two interacting fermions in the presence of their respective Fermi seas.
In our present study, we may therefore approximate~$n_{n_{\uparrow}n_{\downarrow}}$ as follows
\be
n_{n_{\uparrow}n_{\downarrow}}(x_{\uparrow},x_{\downarrow})\approx
N_{\uparrow}N_{\downarrow}\left|\Psi(x_{\uparrow},x_{\downarrow})\right|^2\,,
\ee
which becomes exact in the limit of only one spin-up and one spin-down fermion, i.e. in the absence of the
inert Fermi seas. In any case, an evaluation 
of~$n_{n_{\uparrow}n_{\downarrow}}$ yields the probability to find the spin-up fermion at position~$x_{\uparrow}$ when
the spin-down fermion is located at position~$x_{\downarrow}$.

In addition to the density-density correlation function, the so-called pair correlation function~$n_{\rm pair}$ has attracted a lot of interest
in the search for FFLO-type phases,\footnote{By definition, the pair correlation function is closely related to the 
propagator of the pair. The momentum dependence of the latter has been used in studies of three-dimensional unitary
Fermi gases to detect the onset of inhomogeneous condensation~\cite{Roscher:2015xha}.}
in particular in one-dimensional systems, see, e.g., Ref.~\cite{PhysRevA.78.033607}:
\be
n_{\rm pair}(x,x^{\prime})&=&\langle \psi^{\dagger}_{\uparrow}(x) \psi^{\dagger}_{\downarrow}(x) 
 \psi_{\uparrow}(x^{\prime}) \psi_{\downarrow}(x^{\prime})\rangle\,.
\ee
In terms of the ground-state $N$-body wave function,
this correlation function is given by
\be
&& n_{\rm pair}(x,x^{\prime})\!=\!N_{\uparrow}N_{\downarrow}\!\!\int_{-\infty}^{\infty}\!{\rm d}y_{3}\;\cdots\!\!\int_{-\infty}^{\infty}\! {\rm d}y_{N}\Phi^{\ast}(x,x,y_3,\dots,y_N)\nn\\
&& \qquad\qquad\qquad\qquad\qquad\qquad\times\; \Phi(x^{\prime},x^{\prime},y_3,\dots,y_N)\,.
\ee
In the infinite-volume limit, this function only depends on the difference of~$x$ and~$x^{\prime}$ due to translation
invariance. In this case, its Fourier transform with respect to~$(x-x^{\prime})$, which is nothing but the pair-momentum
distribution, has been found to assume a maximum at 
momenta~$Q_{\rm FFLO}\sim (n_{\uparrow}-n_{\downarrow})\sim (k_{\rm F,\uparrow}-k_{\rm F,\downarrow})$,\footnote{In one dimension, the Fermi momentum~$k_{\rm F,\sigma}$
is proportional to the density~$n_{\sigma}$.} where~$n_{\uparrow,\downarrow}$
is the density of the spin-up and spin-down fermions, respectively.~\cite{PhysRevA.78.033607,Lee:2011zzo}.
Since~$Q_{\rm FFLO}$ is expected to determine
the periodicity of the ground state in the many-body phase diagram~\cite{FuldeFerrell64,LarkinOvchinnikov64}, the existence of a maximum
around~$Q_{\rm FFLO}$ in the pair-momentum distribution
is considered as an indicator for the formation of an inhomogeneous ground state.
{\it A priori}, however, the presence of such a maximum does not necessarily entail that the pairs 
with momenta~$Q_{\rm FFLO}$ describe bound states. Moreover, it does not imply that 
these states are the lowest-lying states in the spectrum and that a condensate is formed from these states.
Still, a pronounced maximum at~$Q\approx Q_{\rm FFLO}$ in this distribution may be viewed as an indication that pairs with momenta~$Q_{\rm FFLO}$ are energetically
most favored and therefore it may serve as an indicator for the formation of a FFLO-type ground state.

In the presence of hard walls, translation invariance is broken explicitly and the pair-momentum distribution  
depends on~$x$ and~$x^{\prime}$ separately. It is therefore convenient to define the auxiliary 
function~$\tilde{n}_{\rm pair}(p,p^{\prime})$,
\be
&&\tilde{n}_{\rm pair}(p,p^{\prime}) \nn\\
&& \qquad = \int_{-\infty}^{\infty}{\rm d}x\int_{-\infty}^{\infty}{\rm d}x^{\prime}\,
n_{\rm pair}(x,x^{\prime})\,{\rm e}^{-{\rm i}(px + p^{\prime}x^{\prime})}\,.
\ee
From this function, we can compute the conventional pair-momentum distribution~$n_{\rm pair}(Q)$:
\be
&& n_{\rm pair}(Q)\nn \\
&& \qquad = \bigintsss_{-\infty}^{\infty} \frac{{\rm d}Q^{\prime}}{2\pi}\,\tilde{n}_{\rm pair}(\tfrac{1}{2}(Q^{\prime}\!+\!2Q),\tfrac{1}{2}(Q^{\prime}\!-\!2Q))\,.
\ee
Indeed, assuming that~$n_{\rm pair}(x,x^{\prime})=n_{\rm pair}(x-x^{\prime})$ in the infinite-volume limit,
we have 
\be
\!\!\!\!\!\!\!\!\!\tilde{n}_{\rm pair}(\tfrac{1}{2}(Q^{\prime}\!+\!2Q),\tfrac{1}{2}(Q^{\prime}\!-\! 2Q)) = (2\pi) n_{\rm pair}(Q)\delta(Q^{\prime})\,.
\ee
In our present study we may consider the following approximation of the
pair correlation function:
\be
n_{\rm pair}(x,x^{\prime}) \approx N_{\uparrow}N_{\downarrow}\Psi^{\ast}(x,x)\Psi(x^{\prime},x^{\prime})\,,\label{eq:pairapprox}
\ee
which again becomes exact in the limit of 
only one spin-up and one spin-down fermion. In the infinite-volume limit, as already mentioned above, 
the wave function~$\Psi$ can be written as a product of a wave function~$\phi_r$ describing
the relative motion of the two fermions in the medium and a wave function for
their center-of-mass motion~\cite{Roscher:2013cma,Roscher:2015xha}:~$\Psi(x_{\uparrow},x_{\downarrow})\propto\phi_{\rm r}(x_{\uparrow}-x_{\downarrow})
\cos(Q_0(x_{\uparrow}+x_{\downarrow})/2)$, where~$Q_0$ is the center-of-mass momentum. 
From this, we obtain
\be
&& \!\!\!\!\! n_{\rm pair}(Q) \nn\\
&& \;\; \sim (2\pi)[\delta(2(Q-Q_0)) + \delta(2(Q+Q_0))] + \text{const.}\,,
\ee
i.e., loosely speaking, the pair-momentum 
distribution in this approximation is sharp at~$Q=\pm Q_0$
in the thermodynamic limit as well.

A word of caution may be required at this point: It 
may very well be just an artifact of the approximation~\eqref{eq:pairapprox}
that we also find~$n_{\rm pair}(Q)$ to be sharp at~$Q=\pm Q_0$. Indeed,
a study of the fully interacting problem in the continuum suggests that~$n_{\rm pair}(Q)$ assumes a  
maximum around~$Q\approx Q_{\rm FFLO}$ even for very small polarizations~$P$~\cite{PhysRevA.78.033607,Lee:2011zzo}. 
Contrary to that, the solution of
the {\it Schr\"odinger} equation~\eqref{eq:Seq} in the thermodynamic limit yields
a distribution that is peaked at~$Q=0$ in the limit of small polarizations~\cite{Roscher:2013cma}. 
Thus, the position~$Q_0$ of the maximum of the momentum distribution~$n_{\rm com}(Q)$ is
in general not identical with the momentum~$Q_{\rm FFLO}$. This is reasonable as all pairs formed in the fully interacting many-problem
contribute to the distribution~$n_{\rm pair}$, whereas~$n_{\rm com}(Q)$ is only a measure for the momentum distribution of possibly formed bound 
states (i.e. states with~$E_{\rm B}>0$). Because of the observed relation of the general structure of the many-body phase diagram and 
the properties of the said bound states in a possibly spin- and mass-imbalanced medium~\cite{Roscher:2013cma,Roscher:2015xha}, we restrict ourselves
to an analysis of the momentum distribution~$n_{\rm com}(Q)$ in the following.

We close this section by emphasizing again that the existence of bound states in our in-medium 
two-body problem does not necessarily entail spontaneous symmetry breaking in the many-body problem. The latter 
requires, additionally, {\it Bose-Einstein} condensation of these bound states. 
In general, a many-body treatment is therefore mandatory in order to obtain the actual phase diagram. However, as 
has been found in previous studies in the thermodynamic limit~\cite{Roscher:2013cma,Roscher:2015xha}, 
the predictions resulting from the {\it Schr\"odinger} equation~\eqref{eq:Seq} for the location of inhomogeneous phases 
turn out to be astonishingly good. This observation emphasizes the importance of few-body physics for our 
understanding of many-body phenomena and sets the stage for our present analysis of a Fermi gas in a hard-wall box.

\section{Results}\label{sec:res}
The main goal of this work is to understand how finite-size effects affect properties of bound states in the 
presence of a mass- and spin-imbalanced 
medium and contrast them with those in the thermodynamic limit.
To this end, it is necessary to disentangle the influence of the various  
``deformations" in our analysis, wherever possible. 
Compared to, e.g., scattering of distinguishable but otherwise equal 
particles in vacuum, which is analytically well understood, the following modifications have to be taken into account:
a finite spatial volume bounded by hard walls, polarization $P=(N_{\uparrow}-N_{\downarrow})/(N_{\uparrow}+N_{\downarrow})$,
mass imbalance~$\bar{m}$, and the fermion densities~$n_{\sigma}=N_{\sigma}/L$. 
In order to relate our in-medium computations in a hard-wall trap with 
previous studies in the thermodynamic limit~\cite{Roscher:2013cma}, we shall keep the overall particle density fixed 
in our present study. This implies that the fermion number increases with increasing~$L$.

Before we now discuss our results in the light of FFLO physics, which requires 
the introduction of either mass or spin imbalance, 
we will first consider the balanced case (including the vacuum limit) and 
characterize finite-size effects that appear already at this stage.
Moreover, we shall discuss the approach to the thermodynamic limit.
The second part is devoted to a discussion of bound state formation in the finite system
and its consequences for the many-body phase diagram.

\subsection{Approaching the infinite-volume limit}\label{subsec:ContLims}

In Ref.~\cite{Roscher:2013cma}, the energies and center-of-mass momenta of bound states 
in the presence of inert Fermi seas have been computed in one dimension in the thermodynamic limit. 
In order to compare to these results and to assess the reliability of the numerical setup underlying 
our present work, 
an analysis of our results in the large-volume limit is in order. 
However, there is a mathematical subtlety that must be taken into account. 
As already discussed above, for any finite volume $L$, 
translational invariance is explicitly broken by the presence of the hard-wall box. 
While the limit $L\rightarrow\infty$ is well defined, it may not 
correspond to a continuous transition to the thermodynamic limit as studied in Ref.~\cite{Roscher:2013cma}.
It is therefore not immediately obvious whether one should indeed expect the finite-box results 
to approach the results from the calculations in the thermodynamic limit as $L\rightarrow\infty$. 
As a truly trapless setup (or periodic boundary conditions) can never be realized in experiments, 
it is important to take {\it a priori} both scenarios into account in our studies.

{\it Vacuum problem.}-- In order to discriminate between finite-size effects and numerical artifacts 
in a controlled way, we will first study two attractively interacting fermions in 
the vacuum, i.e. in the limit of vanishing Fermi seas.
This problem can be solved analytically in the absence of the hard-wall box. 
To be specific, the {\it Schr\"odinger} equation for two particles in vacuum with $\bar{m}=0$ reads
\be
\left[-\sum_{\sigma = \{\uparrow,\downarrow\}} \frac{1}{2m_{\sigma}}\partial_{x_\sigma}^2 
- g\delta(x_\uparrow - x_\downarrow) + E_{\rm B}\right]\Psi(x_\uparrow,x_\downarrow) = 0\,,\nn
\ee
which is nothing but the {\it Schr\"odinger} equation~\eqref{eq:Seq} with~$\epsilon_{{\rm F},\sigma}\to 0$.
In free space, a straightforward solution of this differential equation 
yields a bound state energy of $E_{\rm B} = g^2/8$. 
Recall our sign conventions for~$E_{\rm B}$ detailed below Eq.~\eqref{eq:Seq}.
Note also that we use~$m_{\rm r}=1/4$ for the reduced mass in this work.

In order to extract the binding energy of the two-body problem in vacuum in free space from
our (numerical) solution of the corresponding {\it Schr\"odinger} equation with hard-wall boundary conditions, 
we have to consider a twofold extrapolation scheme: First, for a given value of the box size~$L$, we 
have to consider an extrapolation with respect to the basis size~$N_{\rm B}$, see Eq.~\eqref{eq:PsiExp}. 
The results from this extrapolation
for various values of~$L$ then have to be extrapolated to obtain an estimate for~$E_{\rm B}$ in the
infinite-volume limit.\footnote{A finite value for~$N_{\rm B}$ is associated with an ultraviolet cutoff
for the fluctuations whereas~$L$ represents an infrared cutoff.}
Whereas the volume dependence of binding energies has been studied analytically in the 
literature for boxes with periodic boundary conditions~\cite{Luescher86,Konig:2011ti},
the precise functional dependence of $E_{\rm B}$ on $N_{\rm B}$ and $L$ is unfortunately
not known in a hard-wall box.
Therefore, we consider different types of fit functions. To be specific,
we employ a power-law fit function,
\be
\label{PowerFit}
E_{\rm B}^{(p)}(z) = \frac{\alpha_p}{z^{\beta_p}} + E_{{\rm B},0},\quad z\in\{N_{\rm B},L\}\,,
\ee
an exponential-law fit function
\be
\label{ExpFit}
E_{\rm B}^{(e)}(z) = \alpha_e e^{-\delta_e z^{\beta_e} } + E_{{\rm B},0},\quad z\in\{N_{\rm B},L\}\,,
\ee
and an additional model function for the volume extrapolation 
inspired by {\it Luescher}'s formula for the periodic box~\cite{Luescher86}: 
\be
\label{LuescherFit}
E_{\rm B}^{(l)}(L) = \frac{\alpha_l}{L}e^{-\delta_l L} + E_{{\rm B},0}\,,
\ee
where~$\alpha_p, \alpha_e, \alpha_l, \beta_p, \beta_e, \delta_e, \delta_l$, and~$E_{{\rm B},0}$ are fit parameters.
The differences between the results for~$E_{{\rm B},0}$ from 
the different fit functions may then be used as a measure for the numerical uncertainty
in our determination of the binding energy in the infinite-volume limit. Indeed, in cases where $N_{\rm B}$ and $L$ 
have been chosen sufficiently large in the numerical studies, the choice of the fit function should not affect strongly 
the estimate for the binding energy as the numerical data would essentially be converged in these cases.

In our studies, we can fix the scale by fixing the coupling constant~$g$, or, equivalently, by fixing
the $s$-wave scattering length~$a_s=1/g$. Indeed, we then
have~$E_{\rm B}/g^2 = E_{\rm B}a_s^2 =1/8$. For example, 
choosing~$g=\pi$ (auxiliary units), $gL\in \{\frac{1}{2}\pi^2,\frac{3}{4}\pi^2,\pi^2,\dots,5\pi^2\}$ and~$40\leq N_{\rm B}\leq 155$,
we find
\be
E_{{\rm B},0}/g^2 = 0.1246_{-0.0132}^{+0.0041}\label{eq:2bodyres}
\ee
for~$N_{\rm B}\to\infty$ and~$L\to\infty$. We add that
the infinite-volume limit is found to be approached slowly. For example,
for our largest box size~$gL=L/a_s=5\pi^2$, we obtain~$E_{{\rm B},0}/g^2 \approx 0.1455$ 
from a fit of our data to the ansatz~\eqref{ExpFit}, and~$E_{{\rm B},0}/g^2 \approx 0.1459$ 
when we use Eq.~\eqref{PowerFit}.
The differences in the values for~$E_{{\rm B},0}$ 
from the ans\"atze~\eqref{PowerFit} and~\eqref{ExpFit} are on the sub-percent level 
for the extrapolation to $N_B\rightarrow\infty$. For the infinite-volume extrapolation, however, the 
result for the binding energy~$E_{\rm B}$ from the power-law 
ansatz [lower error bound in Eq.~\eqref{eq:2bodyres}] clearly underestimates the expected value for the binding energy.
On the other hand, the values from the exponential-law ansatz [central value in Eq.~\eqref{eq:2bodyres}] and
the {\it Luescher}-type ansatz [upper error bound in Eq.~\eqref{eq:2bodyres}] agree with the expected analytic value for the binding energy 
up to only a few percent.

Two conclusions may be drawn from this analysis. 
First, a pure power-law ansatz appears to be unsuitable for the description of the volume dependence of the binding energy. 
Of course, this does not come as a surprise: Whereas {\it Luescher}'s formula cannot be expected to hold quantitatively 
for a box with hard walls, it is at least reasonable to assume that the general exponential behavior carries over from the case of periodic 
to our hard-wall boundary conditions. Second, the fits based on the exponential-type as well as
{\it Luescher}-type ansatz yield the analytic result for the binding energy 
within error bars which may very well be traced back to numerical inaccuracy. Recall that two types of extrapolations
are required to obtain the binding energy. In any case, omitting the pure power-law ansatz for the infinite-volume
extrapolation, we find $E_{B,0}/g^2 = 0.1246^{+0.0041}_{-0.0010}$, in very good agreement with the 
exact solution~$E_{\rm B}/g^2=0.125$.  {At least in the vacuum limit, we may therefore state 
that the approach to the (trapless) infinite-volume limit is smooth, in agreement with the analytic solution~\cite{Belloni201425}.}
\begin{figure}
\centering
\includegraphics[width =\columnwidth]{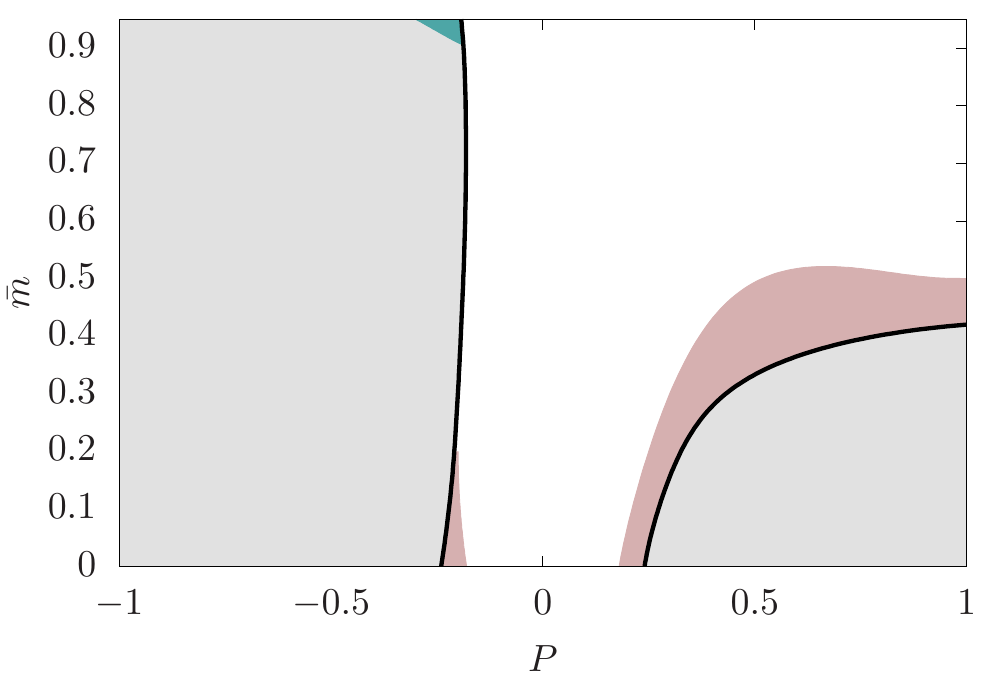}
\caption{Regimes with finite center-of-mass momentum 
in the plane spanned by mass imbalance~$\bar{m}$ and polarization~$P$
as found in the thermodynamic limit (gray-shaded area) and 
in a hard-wall box. For small particle numbers, ``phantom" regimes are found which enlarge (red-shaded areas) 
or diminish (blue-shaded area) the extent of the domain with finite center-of-mass momentum (gray-shaded areas),
see main text for details. The white-shaded area corresponds to a regime where the 
center-of-mass momentum of the bound state is zero.
The various differently shaded 
areas have been obtained from a compilation of the results from the 
box sizes considered in this work.
Note also that the smooth boundaries between the various regimes are only
introduced to guide the eye, as the polarization $P$ is discrete for the finite system. }
\label{fig:PDComp}
\end{figure}

{\it In-medium problem.}-- As discussed above, our in-medium computations are performed at fixed coupling~$g$ and fixed 
overall fermion density in order to allow for a meaningful comparison between systems with different 
box sizes~$L$, including the (trapless) thermodynamic limit~\cite{Roscher:2013cma}. 

The single dimensionless coupling constant in the many-body system is~$\gamma = gL/N = g/n$, 
see also Refs.~\cite{2004PhRvL..93i0405T,2004PhRvL..93i0408F,Rammelmuller:2015zaa}.
This scale fixing implies that the fermion number increases with increasing~$L$. Numerically, 
this poses an additional challenge: As the (inert) Fermi surfaces are shifted to greater particle 
numbers when~$L$ is increased, 
more basis functions are required to ensure the same numerical precision for larger box sizes. 
This is due to the fact that states above and below the Fermi surfaces contribute to the bound-state wave function. 
Note that an increase of the particle number but not $N_{\rm B}$ effectively corresponds to lowering the ultraviolet cutoff.
For our computation of binding energies, we have therefore restricted ourselves to~$N\leq 22$ particles. The density has been
fixed to~$N/L=\pi/2$ (which implies~$\gamma=2$). 
For our maximum number of basis functions $N_{{\rm B},\m{max}} = 155$, the estimates~$E_{{\rm B},0}$ 
for the bound-state energies from the fit functions~\eqref{PowerFit} and~\eqref{ExpFit} for the considered box sizes
then differ only on the sub-percent level. However, whereas the direct computation 
of the bound-state energy in the thermodynamic limit yields $E_{\rm B}/g^2 \approx 0.126$ in the presence of a mass- and
spin-balanced medium, we now extract $E_{{\rm B}}/g^2 = 0.131^{+0.001}_{-0.007}$ from our numerical data 
as~$N_{\rm B}\to\infty$ and~$L\to\infty$.
The lower error bound originates from the extrapolation based on the ansatz~\eqref{PowerFit} 
which we have found to be unreliable in the vacuum case. 
Taking only the exponential-type and {\it Luescher}-type fit ansatz into account for the extrapolation to
the infinite-volume limit, we even find $E_{{\rm B}}/g^2 = 0.131\pm 0.001$. Thus, compared to the value obtained from 
a direct study of the system in the thermodynamic limit, our estimate deviates by about~$4\%$. This suggests
that the convergence of the binding energies to their values in the thermodynamic limit is 
further slowed down in the presence of the Fermi seas. However, within our present setup in terms of 
the sizes of the hard-wall box and the basis, we are not able to clarify whether the thermodynamic limit is
approached continuously, as one may at least naively expect from our two-body study in the vacuum.

\subsection{Phase diagram and finite-size effects}

Up to now, we have only discussed the spin- and mass-
balanced case. Introducing spin- and mass imbalance,
the energetically most favorable center-of-mass momentum $Q_0$ of the lowest-lying bound state becomes a quantity of 
particular interest with respect to a search for inhomogeneous ground states in many-body systems, see our discussion above. 
In order to facilitate the comparison with many-body studies, 
we have computed a "phase diagram"  for the center-of-mass momentum of the
(lowest-lying) bound state as a function of polarization
$P$ and mass imbalance $\bar{m}$. As summarized in Fig.~\ref{fig:PDComp},  
representing our main result,
we find from a numerical solution of the {\it Schr\"odinger} equation~\eqref{eq:Seq} 
that the size of the regimes (``phases") where the lowest-lying bound state carries a finite momentum 
is subject to change with respect to system size~$L$, or, equivalently, with respect to the total particle number~$N$.
Recall that we keep the density fixed.
The black solid lines depict the boundaries of the regime where a finite center-of-mass momentum is energetically favored 
in the thermodynamic limit (gray-shaded area). For a finite system, the size of these ``phases" indeed changes. 
For example, the red-shaded areas in Fig.~\ref{fig:PDComp} correspond to an extension of the gray-shaded areas in the presence of the box. 
These red-shaded areas may be viewed as ``phantom phases" where 
the lowest-lying bound state only carries a finite center-of-mass momentum up to a certain value of the total fermion number.
In the same spirit, the blue-shaded area is a ``phantom phase" associated with a regime where 
a vanishing center-of-mass momentum is preferred at small particle numbers and therefore diminishes 
the gray-shaded area. The white area in Fig.~\ref{fig:PDComp} corresponds to a regime where the center-of-mass
momentum of the bound state is zero. With respect to a study of the many-body phase diagram, the white-shaded
area is associated with a regime where homogeneous condensation is preferred whereas the other three areas 
are associated with regimes where an inhomogeneous ground state may potentially be formed due to the
condensation of bound states with a finite center-of-mass momentum. We add that  
the white-shaded area increases with increasing coupling~$g$, i.e., loosely speaking, homogeneous condensation is favored over
inhomogeneous condensation for increasing coupling strengths. 

In Fig.~\ref{fig:PDComp}, the various areas correspond to a compilation of our results from studies
with different {\it even} total fermion numbers~$N$ ($2\leq N \leq 40$), see below.
We emphasize that 
the polarization~$P$ is discrete for the finite system. Thus, the various differently shaded areas in Fig.~\ref{fig:PDComp}
are not ``continuous" for a given finite system and the smooth ``phase" boundaries for the trapped system are only
introduced to guide the eye. For increasing~$N$, the ``phantom phases" 
shrink and the ``phase diagram" in the thermodynamic limit is approached. 
An astonishing conclusion taken from a comparison of our finite-volume results with 
those from a study in the thermodynamic limit is that the ``phase diagram" in Fig.~\ref{fig:PDComp} is
only mildly altered in the presence of the box, in particular for 
negative polarizations~$P$ and finite mass imbalances~$\bar{m}>0$. As we shall
discuss in detail below, we also observe that the center-of-mass momentum of the bound state
is already close to its value in the thermodynamic limit for~$N\gtrsim 30$ at least in large parts of
the ``phase diagram".

Let us now discuss the finite-size effects underlying the changes of the ``phase diagram" in a hard-wall box as compared
to the thermodynamic limit.
As detailed in Sec.~\ref{sec:form}, we define the center-of-mass momentum 
to be the position~$Q_0$ of the maximum of the distribution~$n_{\rm com}(Q)$, 
see Eq.~\eqref{eq:ncom}. Recall that this is in line with the situation in the
thermodynamic limit, \mbox{where~$n_{\rm com}(Q)\sim \delta(Q-Q_0) + \delta(Q+Q_0)$}.

\begin{figure}
\centering
\includegraphics[width=\columnwidth]{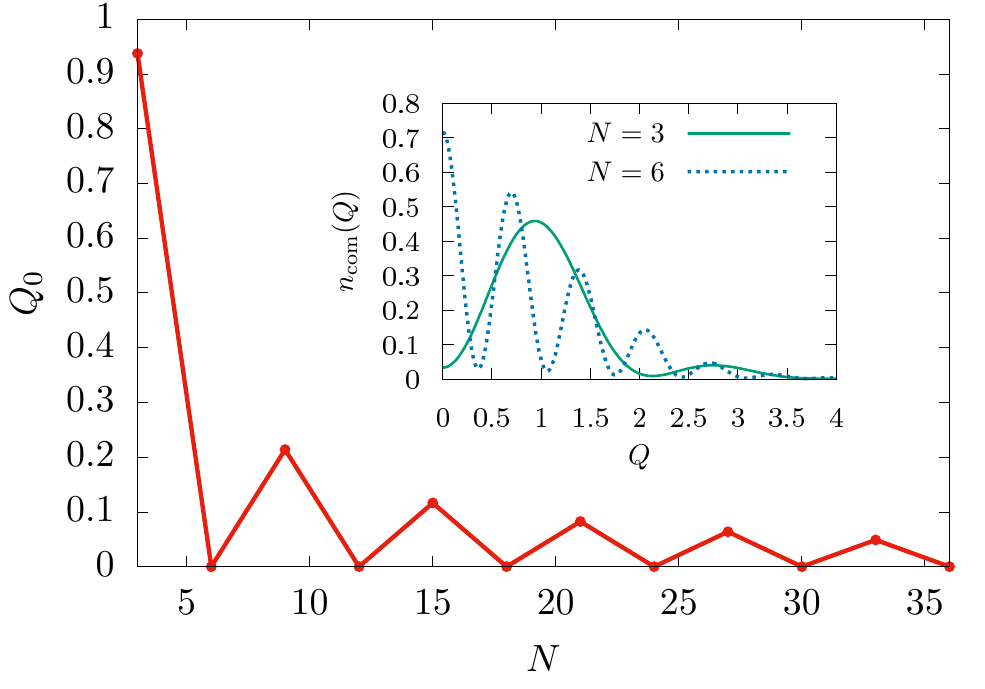}
\caption{Particle-number evolution of the center-of-mass momentum $Q_0$ of the lowest-lying bound state for fixed polarization $P=\frac{1}{3}$ 
and mass imbalance $\bar{m} = 0.8$. In systems with odd total fermion number, the center-of-mass momentum $Q_0$ is found
to be finite and vanishes only as $N\rightarrow\infty$. The inset shows the center-of-mass momentum distribution~$n_{\rm com}(Q)$ 
for~$N=3, 6$.}
\label{fig:UnevenQDownConv}
\end{figure}

For sufficiently large total particle numbers~$N$, it should not matter if $N$ is even or odd. For small~$N$, however, 
we find strong deviations from the results in the thermodynamic limit, 
in particular for odd total particle number. In Fig.~\ref{fig:UnevenQDownConv}, we show~$Q_0$ 
as a function of~$N$ for $\bar{m} = 0.8$ and $P = \frac{1}{3}$. 
As can be read off from Fig.~\ref{fig:PDComp}, $Q_0=0$ should be approached as $L\rightarrow\infty$. This is indeed the case. 
For even~$N$, we observe that~$Q_0=0$ is already assumed for~$N=6$. For odd~$N$, on the other hand,~$Q_0$ is 
only reached as $L\rightarrow\infty$. The reason for this behavior becomes clear when we analyse the functional shape of 
the distribution~$n_{\rm com}(Q)$, see inset of Fig.~\ref{fig:UnevenQDownConv}. For odd~$N$, the symmetries of the 
interaction force the maximum to be located at a finite value of~$Q$.
As this is a physical effect, it suggests that, strictly speaking for any finite 
box size~$L$, there is no regime in the $(\bar{m},P)$ plane where the center-of-mass momentum of the lowest-lying bound state is zero for odd~$N$. 
In particular, for small odd particle
numbers, the center-of-mass momentum is expected to be large.
Therefore, comparisons of the finite-volume phase diagram with the one for the thermodynamic limit 
may be misleading when odd particle numbers are taken into account. 
For this reason, only systems with even $N$ have entered Fig.~\ref{fig:PDComp}.

We add that the appearance of various local maxima in the momentum distribution~$n_{\rm com}(Q)$ 
for both odd and even~$N$ is a generic finite-size effect arising
from the hard-wall boundary conditions.
Therefore it is reasonable to expect that various local maxima are also present in the conventional 
pair-momentum distribution~$n_{\rm pair}(Q)$, in addition to the peak at the FFLO momentum~$Q_{\rm FFLO}$.
This local-maxima effect is most prominent at small~$N$. However, it should not be confused with
the shell structure  
observed in experiments in harmonic traps~\cite{2006PhRvL..97c0401S,PhysRevLett.97.030401,Partridge27012006,PhysRevLett.97.190407,2010Natur.467..567L}.
The latter is also present in the limit of large particle numbers in contrast to the case of a gas in a hard-wall box.

In Sec.~\ref{subsec:ContLims}, we have only discussed fully balanced systems. As the center-of-mass momentum of the lowest-lying bound state
is a key observable but is found to be finite only for imbalanced systems, its convergence properties have to be investigated carefully. 
This is particularly important in view of the extended and diminished finite-$Q_0$ regimes in Fig.~\ref{fig:PDComp}. 
Indeed, for a physical interpretation of the latter to be credible, the possibility of them being merely numerical artifacts must be excluded reliably.

\begin{figure}
\centering
\includegraphics[width=\columnwidth]{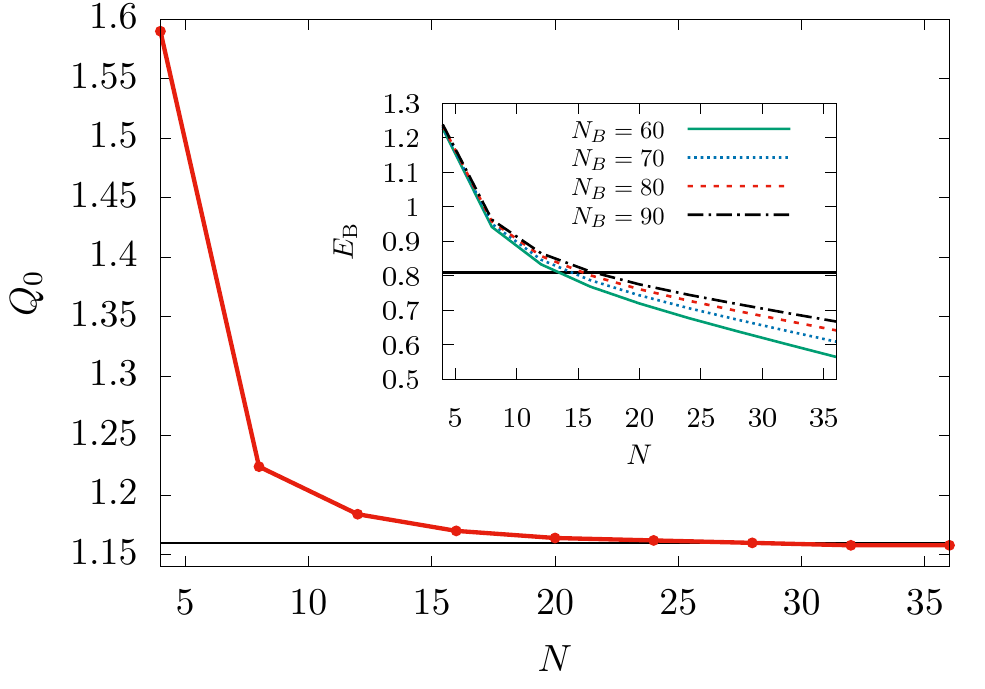}
\caption{Dependence of the center-of-mass momentum $Q_0$ on the total fermion number~$N$ for 
$\bar{m} = 0.5$, $P = -0.5$ and $N_{\rm B} = 60$. Convergence to the result~$Q_0 \approx 1.16$ in the thermodynamic limit (black solid line) 
does effectively not require an extrapolation to the limit~$N_{\rm B}\to\infty$. On the other hand, the corresponding bound-state energy converges
slowly with respect to~$N_{\rm B}$ and~$N$ to its value in the thermodynamic limit,~$E_{{\rm B}} \approx 0.81$, see inset.}
\label{fig:QBulkConv}
\end{figure}

As we are interested in the actual finite-size effects, the extrapolation in terms of $N_{\rm B}$ is the primary source of numerical inaccuracy. 
In order to isolate the latter, let us consider a configuration deep inside a the finite-$Q_0$ regime, e.g. $\bar{m} = 0.5$, $P = -0.5$, see Fig.~\ref{fig:PDComp}.
In Fig.~\ref{fig:QBulkConv}, the corresponding particle-number dependence of $Q_0$ is shown. 
We observe that finite-volume effects are only sizeable for small particle numbers and that
the value in the thermodynamic limit is already reached within sub-percent level accuracy at $N = 30$ in this case, corresponding 
to a system with~$L/a_{s}=2Ng/\pi=2N \gg 1$.
With respect to the basis size~$N_{\rm B}$, it is remarkable that~$N_{\rm B}=60$ already suffices to obtain fully converged results for
the center-of-mass momentum~$Q_0$ for systems with~$N\lesssim 35$. No extrapolation with respect to $N_{\rm B}$ is required. 
As can been seen from the inset of Fig.~\ref{fig:QBulkConv}, however, such small values of $N_{\rm B}$ are by no means sufficient
to obtain a converged estimate for the bound-state energy. In fact, as discussed in Sec.~\ref{subsec:ContLims} for the balanced system, 
basis sizes of~$N_{\rm B}=155$ are at least required to observe convergence in the bound-state energy as a function of~$N_{\rm B}$ for~$N=22$.
Therefore an extrapolation to the limit~$N_{\rm B}\to\infty$ is in general necessary to estimate the binding energy. 
Contrary to that, the center-of-mass momentum~$Q_0$ of the bound
state appears to converge much faster as a function of~$N_{\rm B}$. This may not come as a surprise as 
the bound-state energy is a ``global" property of the wave function and is thus rather sensitive to the cutoff $N_{\rm B}$. On the other hand, 
the position~$Q_0$ of the maximum of $n_{\rm com}(Q)$ is a rather ``local" property in terms of basis functions. 
As long as the global maximum associated with~$Q_0$ 
is sufficiently narrow, its position 
is effectively determined by a small number of basis functions which are then associated with comparatively large coefficients~$c_{kl}$ in Eq.~\eqref{eq:PsiExp}.  
Deep inside a finite-$Q_0$ regime, the global maximum becomes even more pronounced, resulting in the 
observed fast convergence of~$Q_0$ with respect to the cutoff $N_{\rm B}$. When we approach the boundaries 
of the finite-$Q_0$ regimes, the maximum associated with~$Q_0$ is less pronounced and larger values of $N_{\rm B}$ are required to reliably 
determine $Q_0$. In addition to this increased sensitivity on~$N_{\rm B}$, which is noteworthy mainly from a technical point of view, small total particle numbers 
introduce distortions as well. This is particularly true for the red-shaded and blue-shaded ``phantom phases" shown in~Fig.~\ref{fig:PDComp}. 

The origin of the domain where pairing with~$Q_0=0$ is found only for small~$N$ (blue-shaded area in Fig.~\ref{fig:PDComp}) 
can be understood easily. In this region of parameter space, the pair momentum is close to zero even in the thermodynamic
limit. For small system sizes, however, the center-of-mass momentum distribution~$n_{\rm com}(Q)$ exhibits only a few widely separated 
maxima. If $N$ is too small, the distance between the maximum at~$Q=0$ and the first maximum at~$Q\neq 0$  is so 
large that the height of the latter is already smaller than the height of the one at~$Q=0$. 
The value of $Q_0$ in the thermodynamic limit is approached when~$N$ is increased as 
the density of maxima in~$n_{\rm com}(Q)$ increases with $N$, corresponding effectively to an increase of the resolution, 
see also the inset of~Fig.~\ref{fig:UnevenQDownConv}.

The explanation for the existence of the red-shaded ``phantom phase" in Fig.~\ref{fig:PDComp} is slightly more involved: 
Consider first the wave function associated with the energetically lowest-lying state 
of two non-interacting fermions on top of their respective Fermi seas 
without a trapping potential. The center-of-mass momentum of this state corresponds exactly to the difference of the respective 
Fermi momenta. This statement essentially holds also true when the box is present, only the notion of a sharply 
defined center-of-mass momentum is replaced by a more or less prominently peaked distribution $n_{\rm com}(Q)$. 
Thus, any configuration with finite polarization entails a finite center-of-mass momentum in this case. 
Switching on interactions, the situation becomes more complicated since the lowest-lying state of this fermion pair 
may now still favor a vanishing center-of-mass momentum even for a finite imbalance, see Fig.~\ref{fig:PDComp}. Thus, interaction effects 
are to some extent able to counterbalance the effect of an imbalance. Indeed, the actual center-of-mass momentum of the bound state 
is determined by an interplay of the kinetic energy of the two fermions modified by the presence of their Fermi seas 
and the interaction energy. Since we keep the total density~$n$ fixed to the same value for all~$N$, 
the dimensionless interaction strength~$\gamma=gL/N$ is fixed in our studies as well as the Fermi energies associated
with the (inert) Fermi seas. Now the relative weight of the kinetic energy of the two interacting fermions on top of the Fermi seas 
scales as~$\sim 1/L^2= (n/N)^2$ whereas their dimensionless coupling~$\gamma$ is kept fixed when~$N$ is changed.
This suggests that the dynamics of the system is dominated by the kinetic part of the underlying Hamilton operator for small~$N$ which tends to 
favor pairing with finite~$Q_0$ for imbalanced systems. Therefore, the regimes with finite~$Q_0$ are
effectively stabilized in the small~$N$ limit.

\section{Conclusions}\label{sec:conc}

In this work, we have studied finite-size effects on the in-medium bound-state formation of a spin-up and spin-down
fermion in a box with hard walls. In the thermodynamic limit, the properties 
of the potentially formed bound-state have been found 
to serve as a reliable probe for the detection of inhomogeneous 
ground states in the many-body phase diagram spanned by spin imbalance, mass imbalance, 
and also temperature~\cite{Roscher:2013cma,Roscher:2015xha}. 
For such an analysis, the computation of the 
center-of-mass momentum of the bound state is required as it sets the scale for the spatial modulation of the
ground state. Clearly, the bare formation of a bound state with a finite center-of-mass momentum
in a spin- and mass-imbalanced medium does not necessarily entail inhomogeneous
condensation. However, the formation of bound states can be considered as a necessary 
condition for the formation of a condensate. Therefore, in accordance with many-body studies
in the thermodynamic limit~\cite{Roscher:2013cma,Roscher:2015xha}, a study of in-medium bound-state
formation can already be useful to identify regimes in the many-body phase diagram in a hard-wall box where 
FFLO-like inhomogeneous condensation is most likely to occur. 

Our present study of bound-state formation
in a hard-wall box suggests that the question concerning the finiteness of the center-of-mass momenta of the bound state formed in 
a spin- and mass-imbalanced medium is not strongly affected by the presence of the hard-wall box.
In fact, the boundaries between regimes with a finite center-of-mass momentum and 
the one with vanishing center-of-mass momentum appear to be rather insensitive to the presence
of the hard-wall box, in particular for negative polarization and finite mass imbalance, see Fig.~\ref{fig:PDComp}.
Moreover, we observe that the position of the maximum of the momentum distribution of the bound state
converges much faster as a function
of the box size (or, equivalently, total fermion number~$N$) than its energy.
For~$P=-0.5$ and~$\bar{m}=0.5$, for example, the position of the maximum has effectively already assumed
its value in the thermodynamic limit for~$N\gtrsim 30$. This suggests that the value of the center-of-mass momentum in the thermodynamic limit
is approached quickly as a function of the total particle number in one dimension, at least for large 
parts of the phase diagram. Although we have restricted ourselves to one dimension
in our numerical studies, previous studies indicate that our
general observations may very well be carried over to the three-dimensional case~\cite{Roscher:2015xha}.

The actual experimental detection of inhomogeneous phases is challenging. Here,
the measurement of the momentum distribution of bound states may serve as an indicator
for the formation of an inhomogeneous ground state. While the analysis of such distributions
may be promising with respect to the search for inhomogeneous ground states, 
our study also shows that these distributions are affected by the presence of the 
hard walls, at least for small system sizes. This is embodied by the existence of 
clearly separated local maxima in these distributions in addition to the one 
associated with the center-of-mass momentum of the bound state in the thermodynamic limit. 
In experimental studies of momentum distributions aiming at the detection of inhomogeneous
ground states, this has to be taken into account. 

In summary, our results suggest that a hard-wall trap may indeed be worthwhile to consider
as it may serve as an ideal confining potential to bring 
experiment and theory closely together and facilitate the search for exotic inhomogeneous ground states 
in ultracold Fermi gases. Still, our study should only be viewed as a first step
towards an understanding of the dynamics of ultracold Fermi gases in hard-wall boxes. 
A full many-body treatment of spin- and mass-imbalanced Fermi gases, e.g. 
with {\it ab initio} lattice Monte Carlo techniques~\cite{Braun:2014pka,Loheac:2015fxa,McKenney:2015gba},  
beyond our present analysis of bound-state formation
is clearly required to compute the many-body phase diagram of these systems.

{\it Acknowledgments.--~}  The authors would like to thank S.~Diehl, J.~E.~Drut, 
and H.-W.~Hammer  for useful discussions
and collaboration on related projects. 
D.R. acknowledges support by the German Research Foundation (DFG) through the
Institutional Strategy of the University of Cologne within the German Excellence Initiative (ZUK~81) 
and through the Collaborative Research Center SFB~1238, project C04.
J.B. acknowledges support by HIC for FAIR within the LOEWE program of the State of Hesse. In addition, J.B. 
and D.R. also acknowledge support by the DFG under Grant BR \mbox{4005/2-1}. 


%
\bibliography{coldgases}

\end{document}